\documentclass[preprint, amsmath, amssymb, preprintnumbers, showpacs, showkeys, aps, prl]{revtex4-1}
\usepackage{graphicx}
\usepackage{float}
\usepackage{color}

\begin{document}
\preprint{}

\title{Termination dependent topological surface states of the natural superlattice phase Bi$_4$Se$_3$}

\author{Q.D. Gibson$^{1}$}
\author{L.M. Schoop$^{1}$}
\author{A.P. Weber$^{2,3}$}
\author{Huiwen Ji$^1$}
\author{S. Nadj-Perge$^5$}
\author{I.K. Drozdov$^5$}
\author{H. Beidenkopf$^5$}
\author{J.T. Sadowski$^6$}
\author{A. Fedorov$^7$}
\author{A. Yazdani$^5$}
\author{T. Valla$^4$}
\author{R.J. Cava$^1$}
\email{rcava@princeton.edu}
\affiliation{$^1$Department of Chemistry, Princeton University, Princeton, NJ 08544, USA}
\affiliation{$^2$National Synchrotron Light Source, Brookhaven National Lab, Upton, NY 11973, USA}
\affiliation{$^3$Department of Physics and Astronomy, University of Missouri-Kansas City, Kansas City, MO 64110, USA}
\affiliation{$^4$Condensed Matter Physics and Materials Science Department, Brookhaven National Lab, Upton, NY 19973, USA}
\affiliation{$^5$Department of Physics, Princeton University, Princeton, NJ 08544, USA}
\affiliation{$^6$Center for Functional Nanomaterials, Brookhaven National Lab, Upton, NY 11973}
\affiliation{$^7$Advanced Light Source, Lawrence Berkeley National Laboratory, Berkeley, CA 94720}

\date{\today}

\begin{abstract}

   We describe the topological surface states of Bi$_4$Se$_3$, a compound in the infinitely adaptive Bi$_2$-Bi$_2$Se$_3$ natural superlattice phase series, determined by a combination of experimental and theoretical methods. Two observable cleavage surfaces, terminating at Bi or Se, are characterized by angle resolved photoelectron spectroscopy and scanning tunneling microscopy, and modeled by ab-initio density functional theory calculations. Topological surface states are observed on both surfaces, but with markedly different dispersions and Kramers point energies. Bi$_4$Se$_3$ therefore represents the only known compound with different topological states on differently  terminated easily distinguished and stable surfaces.
 \end{abstract}

\maketitle


Three-dimensional topological insulators (3D TIs) are a new class of materials that exhibit topologically 
protected helical metallic topological surface states (TSS) and a bulk band gap \cite{fu2007topological, lin2010single, neupane2012topological, hsieh2008topological, noh2008spin, hsieh2009tunable, 
hsieh2009observation, ShenBi2Te3, ShenTlBiTe2, ZhangBi2Se3, AndoTlBiSe2, AndoBTS}. The great interest in 3D TIs is partly due to the fact that they necessarily host exotic bound states at their boundaries when interfaced with other non-topological or topological materials \cite{Takahashi2011}. In addition, several theoretical studies have predicted that novel properties may emerge when topological insulators are interlaced with other materials in a regular superlattice \cite{Burkov2011,Jin2012}. In order to pursue these promising avenues, however, various experimental challenges have to be solved. For example, a basic understanding of the surface band structures of more complex topological materials, while highly desirable, is not trivial. Although there have been studies of different materials at the interface of 3D TIs\cite{Miao19022013,hirahara2011interfacing}, and on different surface terminations of the same material\cite{PhysRevLett.110.156101,doi:10.1021/nl304099x} no in depth study of the TSS and electronic band structure of a complex topological material, or a true bulk topological superlattice material has yet been reported.         

Here we investigate the properties of Bi$_4$Se$_3$, the simplest  topological superlattice material, consisting of single Bi$_2$ layers interleaved with single Bi$_2$Se$_3$ layers in a 1:1 ratio\cite{bayliss1991crystal}(Fig 1(a).  While bulk Bi$_2$Se$_3$ is a model 3D TI, an isolated Bi$_2$ layer is predicted to be a 2D TI \cite{Murakami2006}; combining these two building blocks into a 3D superlattice offers a unique possibility for studying the effects of inter-layer interactions. We investigate the electronic structure of this material experimentally via angle-resolved photoemission spectroscopy (ARPES) and scanning tunneling microscopy (STM) and theoretically via ab-initio density functional theory (DFT) calculations. 
We observe two types of surfaces after cleaving the crystal, corresponding to Bi$_2$ and Bi$_2$Se$_3$ terminated terraces. We find that both terminations exhibit TSS, but with substantially different Kramers point energies and dispersions. We show that many features of the surface band structure can be derived from the idealized case of weakly coupled Bi$_2$ and Bi$_2$Se$_3$ layers where the interaction between these building blocks is responsible for the different TSSs.  Bi$_4$Se$_3$ and related( Bi$_2$)$_m$(Bi$_2$Se$_3$)$_n$ \cite{bos2007structures} superlattice phases provide a unique opportunity for studying the coexistence of multiple types of topological surface states on distinct, stable, and separable cleavage surfaces in the same material.

Crystals of Bi$_4$Se$_3$ were grown by slow cooling a Bi-rich melt. The crystal structure and quality were confirmed by X-ray diffraction. The ARPES experiments were carried out on a Scienta SES-100 electron spectrometer at beamline 12.0.1 of the Advanced Light Source. The spectra were recorded at photon energies ranging from 35 to 100 eV, with a combined instrumental energy resolution of $\sim$15 meV and an angular resolution better than $\pm 0.07^{\circ}$. The combined spatial resolution, dependent upon precise linear motion control of the sample and the 60 $\mu m$ photoemission spot size of the beam, was better than 80 $\mu m$. Samples were cleaved at 15-20 K under ultra-high vacuum (UHV) conditions. The temperature was measured using a silicon sensor mounted near the sample.
Photoemission Electron Microscopy (PEEM) experiments were carried out at the XPEEM/LEEM end-station at the National Synchrotron Light Source beamline U5UA at room temperature with 47.8 eV photons using an Elmitec SPELEEM III microscope. All the samples were cut from the same bulk piece and cleaved and measured in ultrahigh vacuum conditions (base pressure better than 2 x 10$^{-9}$ Pa in the ARPES chamber and better than 2 x 10$^{-8}$ Pa in the PEEM chamber). Samples for STM measurements were cleaved {\it in-situ} at room temperature under ultra-high vacuum conditions, with the measurements performed at 4.2 K. Both Bi$_2$ and Bi$_2$Se$_3$ surfaces are exposed in the cleaves; there is no known dependence of terrace size on cleavage temperature.
   
  Surface electronic structure calculations were performed in the framework of density functional theory using the Wien2k code \cite{Blaha1990399} with a full-potential linearized augmented plane-wave and local orbitals basis together with the Perdew-Burke-Ernzerhof parameterization of the generalized gradient approximation \cite{perdew1996generalized}, using a slab geometry. The plane wave cutoff parameter R$_{MT}$K$_{max}$ was set to 7 and the Brilloun zone (BZ) was sampled by 9 k-points, or 100 k-points in the case of the weakly coupled Bi$_2$ and Bi$_2$Se$_3$ layers. Spin-orbit coupling (SOC) was included. For the Bi$_2$ terminated surface, a slab was constructed of 5 Bi$_2$Se$_3$ layers and 6 Bi$_2$ layers, with 10~{\AA} of vacuum between adjacent slabs. For the Bi$_2$Se$_3$ terminated surface, a slab was constructed of 6 Bi$_2$Se$_3$ layers and 5 Bi$_2$ layers. To calculate weakly coupled Bi$_2$ and Bi$_2$Se$_3$ layers, a rhombohedral unit cell was used that retains all of the parameters of Bi$_4$Se$_3$ but with an artificial interlayer distance of 5~{\AA}. The experimentally determined lattice parameters and atom positions were used to construct the slabs. The contribution of the surface atoms to the overall surface electronic structure was determined by calculating the partial contribution of each atomic basis set to the wavefunctions at all k-points.

Bi and Se rich regions on cleaved surfaces, corresponding to Bi$_2$ bilayers and Bi$_2$Se$_3$ quintuple layers, were identified in PEEM by difference in work function (not shown). Subsequently, micro-spot x-ray photoemission spectra (micro-XPS) for the Bi 5d core level(Fig. 1(b)) were obtained from Bi$_2$ and Bi$_2$Se$_3$ regions, respectively. The spectra taken on the Bi$_2$Se$_3$ are shifted by about 1.8eV towards higher binding energies for both components of the Bi 5d doublet. We further utilized the respective Bi 5d 5/2 component from both regions to obtain PEEM images of the surface(Fig. 1(c)). The terraces with different terminations are clearly visible, ranging in size from a few $\mu m^2$ to $\sim (100\mu m)^2$, The fine features in the topography measured in STM clearly reveal two types of surfaces (Figs 1(d) and 1(e). The topography of the region close to the step edge allows us to identify the surfaces; the measured step heights are approximately 4 {\AA} and 8 {\AA}, consistent with the first and the second surfaces being a Bi$_2$ bilayer and Bi$_2$Se$_3$ quintuple layers respectively. Domains suitable for quasiparticle interference (QPI) measurements were observed.

While the~60 $\mu m$ ARPES spot size is comparable to the domain size of the different terminations, and thus insufficient to completely resolve the two terminations, Bi- and Se-predominated surface terminations are easily distinguished, allowing for the clear determination of the surface electronic structures of both terminations. Figure 2(a) shows the ARPES spectra of the Se-rich and Bi-rich surfaces, reflecting significant differences between the two terminations. The Bi 5d and Se 3d core levels in Fig. 2(b) were measured at exactly the same locations where the ARPES spectra from Fig. 2(a) were recorded, enabling the identification of two different surface states with two terminations.
   
   Contrary to expectations, the Bi$_2$ termination exhibits nearly linear Dirac surface states similar to those observed in topological insulators such as Bi$_2$Se$_3$, although with the hole-like dispersion, while the Bi$_2$Se$_3$ termination exhibits a non-linearly dispersing surface state. The existence of these surface states is consistent with the band inversion at $\Gamma$ for bulk Bi$_4$Se$_3$ (indicating a nontrival Z2 invariant) and the with the TSS observed earlier on Bi$_4$Se$_{2.4}$S$_{0.6}$ \cite{valla2012topological}.
         
   The calculation for the Bi$_2$Se$_3$ termination (left, Fig. 2(c)) shows the same type of non-linear surface state that is experimentally observed by ARPES. The calculated electronic structure indicates that the  state is non-trivial; there is a symmetry-allowed crossing at about -0.5 eV and the state crosses the continuous gap in the bulk bands an odd number of times along $\Gamma$-M, therefore satisfying the odd-crossing criterion for a topological surface state. The state is similar to those states seen on the surface of topological elemental Sb \cite{hsieh2009observation2}.     
   
    The calculated surface states for the Bi$_2$ termination (right, Figure 2(c)) clearly show a surface Dirac cone. A close look reveals that the surface contribution vanishes precipitously upon joining the semimetallic bulk bands slightly above E$_F$, meanining that the surface state  crossing the continuous gap an odd number of times. The observed surface electronic structure bears resemblance to that found on single Bi$_2$ layers deposited on Bi$_2$Te$_3$, which has been suggested as a platform for quantum spin hall (QSH) edge states \cite{hirahara2011interfacing}.  A small continuous gap in the calculated surface state spectrum is observed at about 0.5 eV above E$_F$. 
       
	Comparisons of the observed and calculated surface electronic structures of the Bi$_2$ and Bi$_2$Se$_3$ terminated surfaces of Bi$_4$Se$_3$ are shown in Figure 2(d), displaying a good match between calculation and experiment. The real crystal is slightly n-doped (by about 0.1 eV) when compared to the calculations.

  To explain the overall surface state electronic structure of Bi$_4$Se$_3$, the calculated electronic structure was compared to that of weakly coupled Bi$_2$ and Bi$_2$Se$_3$ layers (separated by 5 {\AA})(Fig. 3(a)). The weak coupling was confirmed by a lack of dispersion along Z. In the completely uncoupled case, QSH  edge states on the single Bi$_2$ layer would lie in the gap above E$_F$ between the valence and conduction bands of the Bi$_2$ layer. In the weak coupling case, there is a distinct band inversion between the conduction band (lCB1$_{Bi_2Se_3}$) of Bi$_2$Se$_3$ and the valence bands (VB1$_{Bi_2}$ and VB2$_{Bi_2}$) of Bi$_2$, with a small SOC induced gap (about 20 meV) appearing between VB1$_{Bi_2}$ and CB1$_{Bi_2Se_3}$ around E$_F$. This is a topological band inversion; the parity of the VB1$_{Bi_2}$ is opposite that of both CB1$_{Bi_2Se_3}$ and VB2$_{Bi_2}$. There is also an avoided crossing gap(about 200 meV) between VB2$_{Bi_2}$ and CB1$_{Bi_2Se_3}$. The Kramers point energies and dispersions  of the Bi$_2$ and Bi$_2$Se$_3$ terminated TSS correspond extremely closely to the VB2$_{Bi_2}$ maximum and CB1$_{Bi_2Se_3}$ minimum, respectively, of the weakly coupled case. This suggests that the Kramers point energies of the TSS are determined by energy levels intrinsic to the individual, isolated version of the layer that hosts them. 

 Figure 3 (b) shows the corresponding Bi$_2$ derived features in the more realistic surface band structure that includes the full interlayer coupling.  Full coupling between the layers moves  CB1$_{Bi_2Se_3}$ up in energy and the valence bands of Bi$_2$ down, retaining the band inversion at $\Gamma$ but un-inverting the bands at Z in the bulk electronic structure, allowing for a nontrivial Z2 invariant. Full coupling also increases the SOC gap between CB1$_{Bi_2Se_3}$ and VB1$_{Bi_2}$ , and increases the avoided crossing between CB1$_{Bi_2Se_3}$ and VB2$_{Bi_2}$ to about 0.6eV. The small gap at 0.5 eV above E$_F$ is between bands mainly derived from the Bi$_2$ and Bi$_2$Se$_3$ conduction bands and Bi$_2$ valence bands. However, all of the bands are heavily hybridized with each other; it would be interesting to see if QSH states could be hosted in that gap, as all of the involved bands have significant Bi$_2$ character.

The raw ARPES data, which does not reflect the full 6-fold symmetry of the system, was 6-fold symmetrized in order to compare to the DFT calculations and STM data (Fig.4). Comparison between the ARPES data and DFT calculations then gives insight to the electronic structure around the Fermi energy, and allows for the construction of schematic CECs (Fig. 4(c) and (f)) and the identification of observed bands. We note A close look reveals that the surface contribution vanishes precipitously upon joining the semimetallic bulk bands slightly above E$_F$. that the high-k feature at both  E$_F$ and -0.75 eV, while derived from the Bi$_2$ terminated surface states, is not related to the TSS and is likely a trivial surface state.

To investigate the local spectroscopic signatures of the two surfaces and make a connection to the ARPES data we performed STM of the interference patterns caused by surface defects (Fig.5). In the range of -100 mV to +100 mV we observe QPI patterns on both Bi$_2$ and Bi$_2$Se$_3$ terminated surfaces. The Fourier transform of the real space conductance map can be linked directly to the joint density of states(JDOS) calculated using ARPES data (see Fig. 5(c) and Fig. 5(f)). Importantly, on both surfaces we observe an overall suppression of the scattering intensity. The observed suppression is at least partly due to backscattering protection coming from the spin texture, as in the case of usual topological insulator surfaces \cite{roushan2009}. This is consistent with 
previous spin-resolved ARPES experiments on Bi$_4$Se$_{2.4}$S$_{0.6}$\cite{valla2012topological}. The only well resolved scattering vectors are the ones along the $\bar{\Gamma} $ - $\bar{M}$ direction, marked by blue and green lines in Fig. 5(a). These vectors can be easily identified as the scattering from the inner ring-like parts of the Bi$_2$ surface bands close to the outer bands which are not protected by spin texture (see also Fig. 4(c)). In order to make quantitative comparison between STM data and JDOS details of the hybridization between bulk and surface bands have to 
be taken into account. This may be the subject of future study.

In conclusion, we have shown that TSS are observed on both types of cleaved surfaces of the natural superlattice phase Bi$_4$Se$_3$. The dispersion and Kramers point energy of the TSS are shown to differ between the two surface terminations, and from simple expectations. This provides the first example of distinct TSS on different surfaces with the same crystallographic orientation in a complex material. The observations and analysis show that the electronic features of a Bi$_2$ single layer are present in Bi$_4$Se$_3$, implying that a QSH state may be hosted in the gap 0.5 eV above E$_F$. The QSH state is potentially observable by STM experiments at a simple step-edge on the surface of Bi$_4$Se$_3$, which would expose the edge of a single Bi$_2$ layer. If present, the existence of QSH states in Bi$_4$Se$_3$ would provide a unique opportunity for studying the coexistence of 1D and 2D topological electronic states in a bulk single crystal. Finally, we show that the difference in TSS in Bi$_4$Se$_3$ is due to the interaction of the building blocks; this suggests that modification of the TSS on the surfaces of topological materials may be experimentally realizable in the large family of natural superlattice materials in the Bi$_2$-Bi$_2$Se$_3$, Bi$_2$-Bi$_2$Te$_3$ and Sb$_2$-Sb$_2$Te$_3$ systems.

\bigskip 
\begin{acknowledgments}

The authors acknowledge helpful discussions with B.A.Bernevig and F.Chen. The financial support of the National Science foundation, grants NSF-DMR-0819860 and NSF-DMR-1104612, DARPA-SPAWAR grant N6601-11-1-4110 and the ARO MURI program, grant W911NF-12-1-0461, are gratefully acknowledged. The work at Brookhaven was carried out in part at the Center for Functional Nanomaterials and the National Synchrotron Light Source which are supported by the US Department of Energy (DOE), Office of Basic Energy Sciences, under Contract No. DE-AC02-98CH10886. The Advanced Light Source is supported by the U.S. DOE, Office of Basic Energy Sciences, under Contract No. DE-AC02-05CH11231. One of the authors (S. N-P) acknowledges support of European Community through the Marie-Curie OEF fellowship.

\end{acknowledgments}

\bibliography{Bi4Se3_paper_bibliography}

\begin{figure}[h]
  \centering
  \includegraphics[width=15cm]{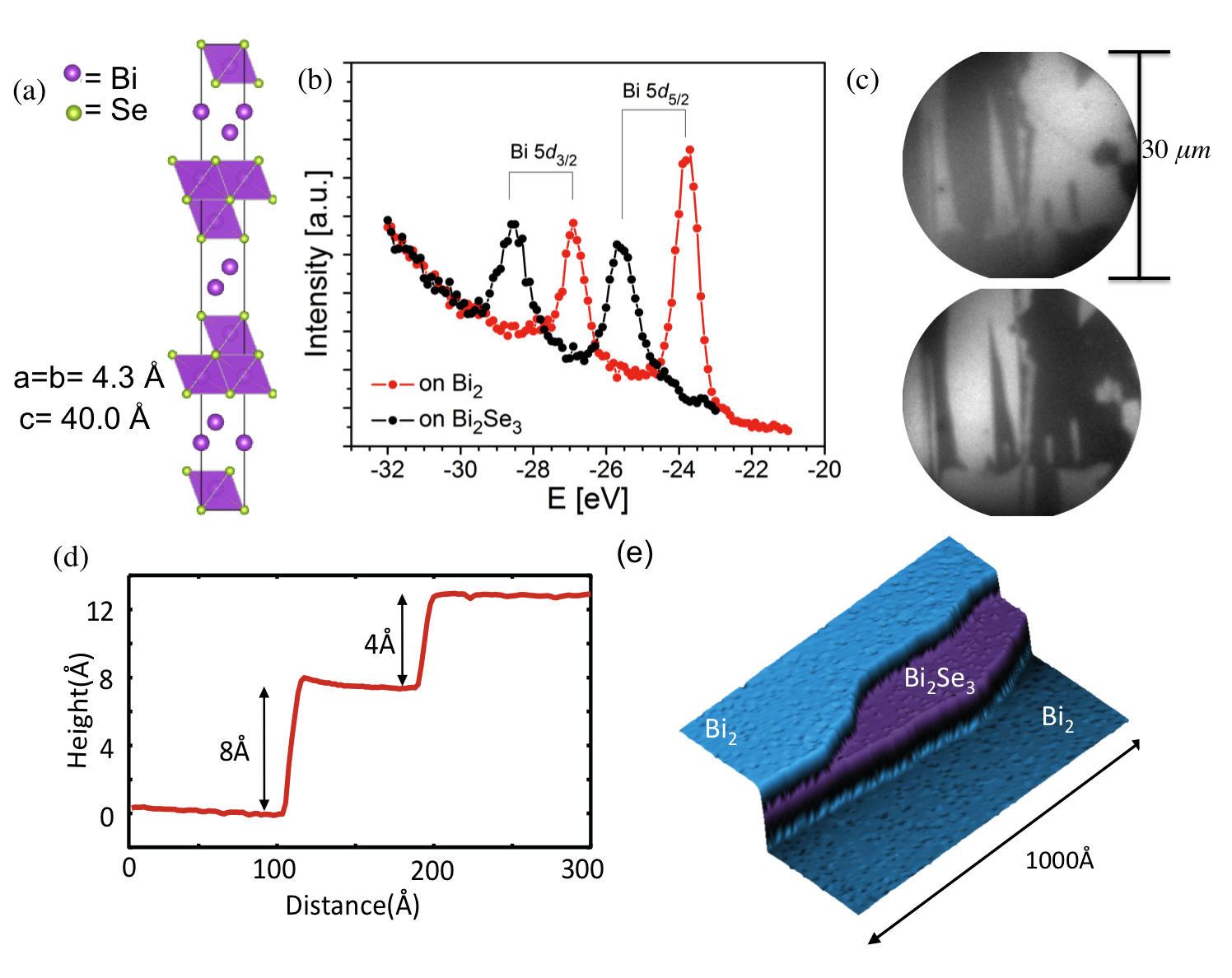}
  \caption{(Color Online)(a) Unit cell of Bi$_4$Se$_3$ projected down the a axis. Bi-Se polyhedra are shaded to highlight the alternating quintuple layer-bilayer structure. 
 (b)  Micro-XPS spectra for the Bi 5d core level, taken from Bi2 and Bi2Se3 regions, respectively; spot size 2 micrometers.
  (c) PEEM images obtained using respective Bi 5d 5/2 core levels  showing high intensity (bright) for the Se termination (top) and for the Bi termination (bottom). The field of view is 30 micrometers. The average domain size of the different terminations are approximately equivalent.
  (d) STM Line-cut across a step edge showing the heights of the Bi$_2$ and Bi$_2$Se$_3$ steps (4 {\AA} and 8 {\AA} respectively).
(e)False color STM topography image close to the step edge from d) where both types of surfaces can be identified}
  \label{fig_1}
\end{figure}

\begin{figure}[h]
  \centering
  \includegraphics[width=15cm]{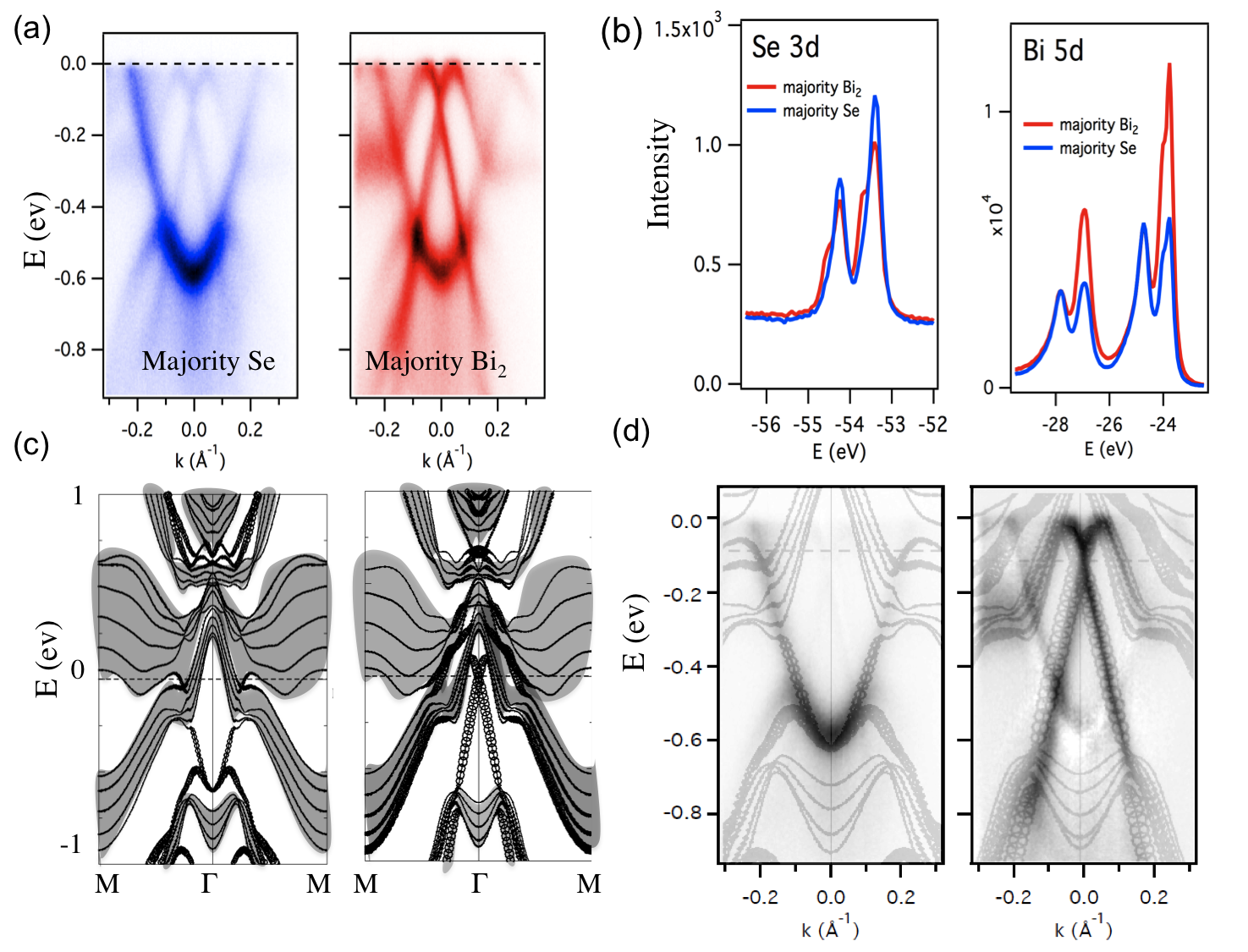}
  \caption{(Color Online)(a) Raw ARPES data for cleavage surfaces dominated by Se termination (left) and Bi termination (right).  (b) Photoemission spectra of Se 3d and Bi 3d levels used to identify Bi$_2$Se$_3$ and Bi$_2$ rich regions. (c) Calculated surface electronic structure for the Bi$_2$Se$_3$ terminated (left) and Bi$_2$ terminated (right) surfaces of Bi$_4$Se$_3$. Bulk bands are shaded. (d) Calculated surface electronic structures of Bi$_2$Se$_3$ (left) and Bi$_2$(right) terminated surfaces, overlaid on the surface states observed experimentally by ARPES, on the same scale. }
\label{fig_2}
\end{figure}
 \begin{figure}[h]
  \centering
  \includegraphics[width=15cm]{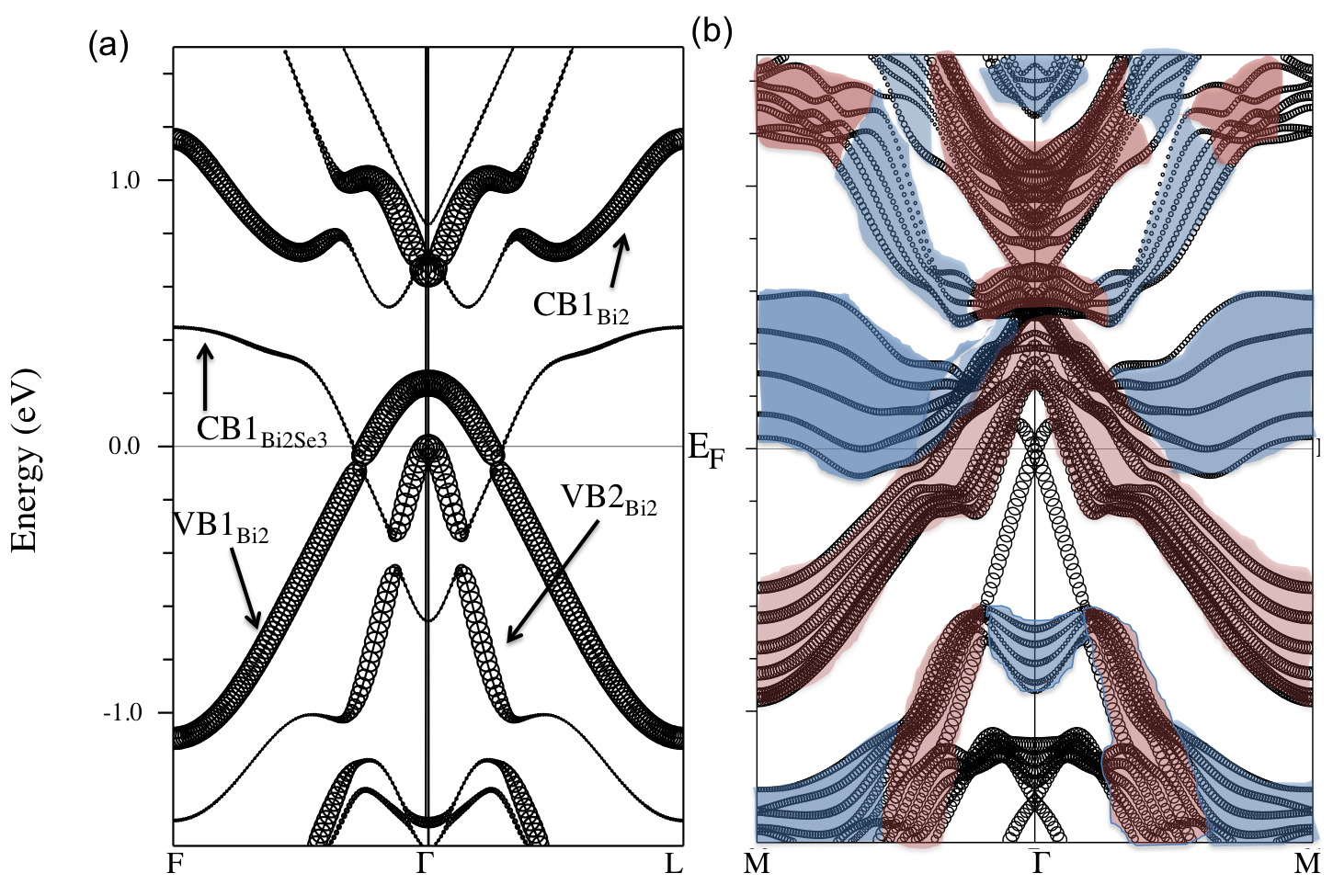}
  \caption {(Color Online)(a) The calculated electronic structure of weakly coupled Bi$_2$ and Bi$_2$Se$_3$ layers. Both F and L would project to $\bar{M}$ in the surface electronic structure. (b) The calculated surface state electronic structure for the Bi$_2$ terminated surface of Bi$_4$Se$_3$, shown for comparison to the idealized case. Heavier plotting shows the contribution of the Bi$_2$ layers. Bulk bands derived mainly from Bi$_2$ and Bi$_2$Se$_3$ are shaded red and blue, respectively. For both (a) and (b) circle size is proportional to the amount of Bi$_2$ bilayer character .}
\label{fig_3}
\end{figure}
\begin{figure}[h]
  \centering
  \includegraphics[width=15cm]{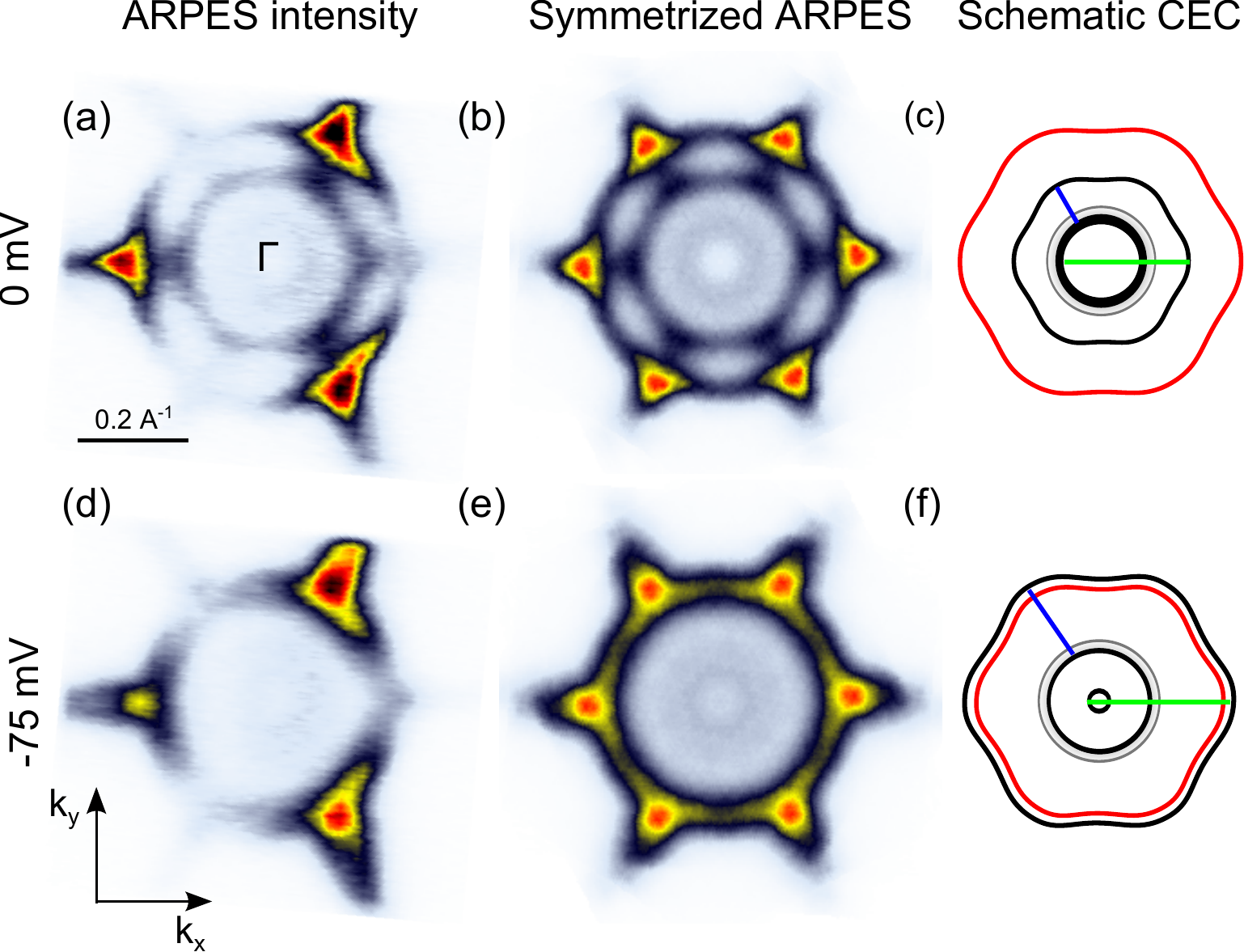}
  \caption{(Color Online)(a) and (d) The ARPES intensity slices corresponding to the Fermi surface (E = 0 meV) and E = -75 meV, respectively. Directions $\bar{\Gamma}-\bar{M}$ and $\bar{\Gamma}-\bar{K}$ are along $k_x$ and $k_y$, respectively. (b) and (e) Six fold symmetrized ARPES data for E = 0 meV and E = -75 meV. (c) and (f) Schematic CECs constructed from comparing the ARPES data with the calculations, for the Bi$_2$ termination. Red, black and gray lines are from the Bi$_2$Se$_3$ surface states, Bi$_2$ surface states and bulk bands, respectively. Blue and green lines mark the dominant scattering vectors observed in STM.  (see Fig. 5).}
  \label{fig_4}
\end{figure}

\begin{figure}[h]
  \centering
  \includegraphics[width=15cm]{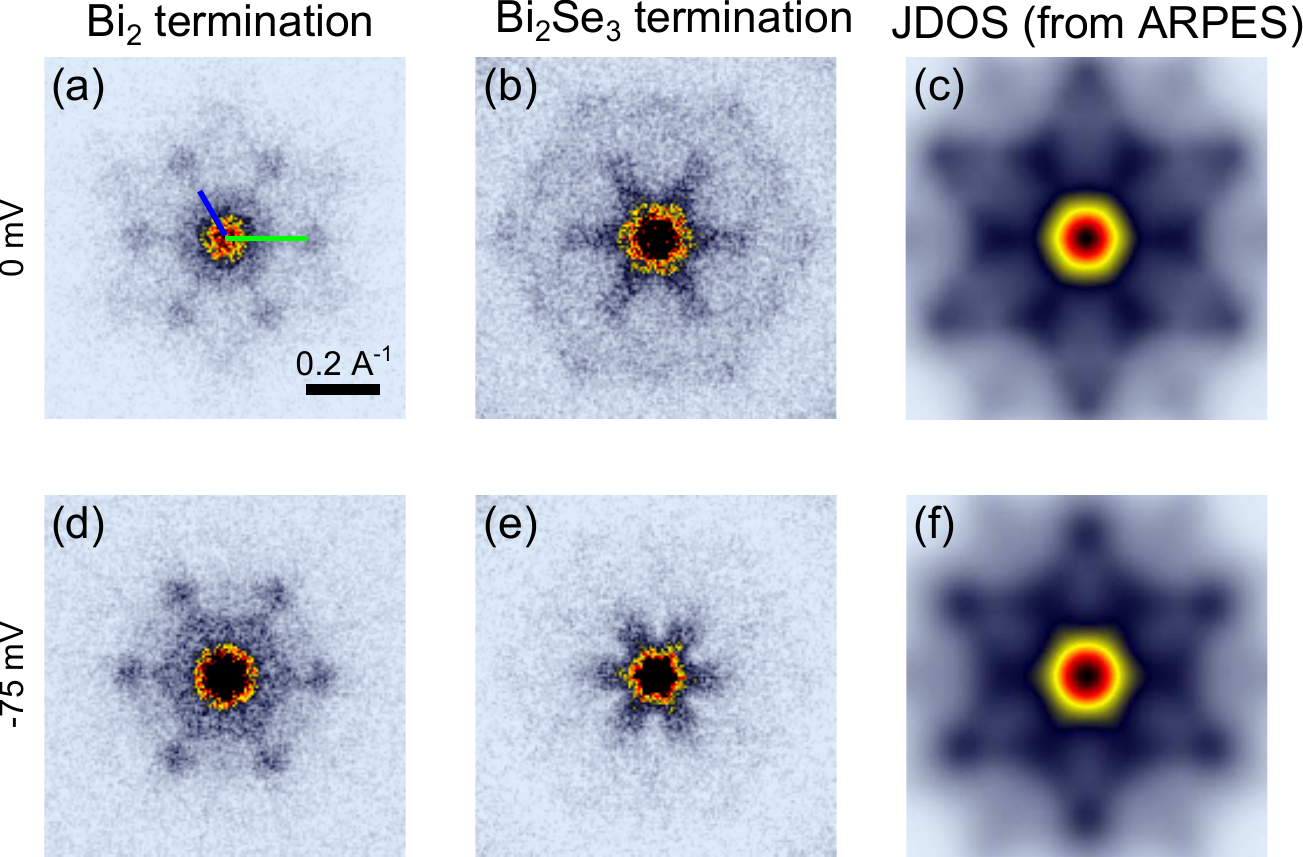}
  \caption{(Color Online)(a) and (d) Fourier transform of the STM conductance map on the Bi$_{2}$ terminated surface for V = 0 mV and V = -75 mV .The color map shows intensities from white (low) to red-black (high). The scattering vectors corresponding to the inner surface structure of the Bi bilayer to the outer bands along  the $\bar{\Gamma}-\bar{M}$ direction is marked by the blue and green lines. (b) and (e) Fourier transform of the STM conductance for the same voltages on the Bi$_2$Se$_3$ surface. (c) and (f) Joint density of states calculated using ARPES data corresponding to the Fermi surface and E=-75 meV.}
  \label{fig_5}
\end{figure}

\end{document}